\documentclass[12pt]{article}
\usepackage{graphicx}
\title{Model-independent  view on the low-mass proton-antiproton enhancement
\\}
\author{B.Kerbikov, A.Stavinsky, V.Fedotov\\
 State Research
Center\\Institute of Theoretical and Experimental Physics, \\
Moscow, Russia}
 \date{}
  \newcommand{\be}{\begin{equation}}
\newcommand{\ee}{\end{equation}}
 
\def\fun#1#2{\lower3.6pt\vbox{\baselineskip0pt\lineskip.9pt
\ialign{$\mathsurround=0pt#1\hfil ##\hfil$\crcr#2\crcr\sim\crcr}}}

\newcommand{\vep}{\mbox{\boldmath${\rm p}$}}

\begin{document}
\maketitle

\begin{abstract}
We present a simple interpretation of  the recently observed
near-threshold proton-antiproton enhancement. It is described by a
set of low-energy parameters deduced from the analysis of $N\bar
N$  experiments at LEAR.
 We predict a
related effect in photoproduction reaction under study by CLAS
collaboration.
\end{abstract}

%\section{Introduction}

Low-mass baryon-antibaryon enhancement has recently been observed
in the decays $B^+\to K^+p\bar p$ \cite{1}, $\bar B^0 \to D^0p\bar
p $ \cite{2} and $J/\psi \to \gamma p\bar p$ \cite{3}. On the
theoretical side this discovery has been discussed by several
authors   \cite{4,5,6,7}. In \cite{4} the near-threshold effect has
been considered either as a gluonic state or as a result of the
quark fragmentation process. In \cite{5} it has been attributed to
the Breit colorspin interaction while in \cite{6} it has been
regarded as caused by peripheral  one-pion-exchange potential. The 
quantum numbers corresponding to the observed enhancement are 
discussed in \cite{7}.

The purpose of the present work is to show that the recently
observed near-threshold $p\bar p$ enchancement may be understood
invoking the knowledge on $N\bar N$ interaction gained at the Low
Energy Antiproton Ring (LEAR) at CERN. The LEAR results may be
schematically subdivided into three groups:

(i) meson  spectroscopy,

(ii) baryonium  searches,

(iii) $N\bar N$ interaction: scattering, annihilation, protonium
spectroscopy.

A comprehensive review of the investigations performed at LEAR
along these lines may  be found e.g. in \cite{8}. At first sight
it might seem tempting to connect the newly observed structure
with the elusive baryonium \cite{9}-\cite{11}. The point of view
adopted in this paper is different. We remind that baryonium
searches at LEAR ended without any clear evidence for its
existence \cite{8}. On the other hand the studies of  $N\bar N$
elastic scattering, annihilation and  charge-exchange reaction
along with the protonium studies yielded reliable information on
low-energy $N\bar N$  amplitude \cite{8}. It has been shown
\cite{12}-\cite{15} that the whole set of the experimental data on
low-energy $N\bar N$ interaction is possible to describe in terms
of the effective range approximation. In this approach the lack of
the dynamical picture is traded for the possibility to reconcile
within a unique scheme different pieces of information on $N\bar
N$ interaction.

We shall see  that the effective range analysis
\cite{12}-\cite{15} based on the old LEAR data enables to explain
Belle \cite{1,2} and BES \cite{3} results. In particular the
observation by BES of a strong effect in the decay $J/\psi\to
\gamma p \bar p$ and the absence of a similar structure in $J/\psi
\to \pi^0 p\bar p$ perfectly fits into the solution for the
low-energy parameters obtained 15 years ago \cite{12}-\cite{15}.
We shall return to this point below.

The problem of whether any  conclusions  on baryonium  can be
inferred from the low-energy parameters is beyond the scope of
this work. We only mention that the values of the  scattering
lengths presented below are not anomalously large as it should be
for the case of a level close to threshold. One must also keep in
mind that strong annihilation  drastically changes the standard
relation between the low-energy parameters and the positions of
the poles of the amplitude \cite{16}.

Our approach is based on the low-energy analysis of final-state
$p\bar p$ interaction. Therefore the near-threshold enhancement
observed in different reactions \cite{1,2} is described  by the
same equations presented below. Another  consequence of the
proposed scheme is the prediction of the similar phenomena in other
reactions of the  same type, in  particular in photoproduction
$\gamma p \to pp\bar p $.
 This process can be investigated by CLAS
collaboration at Jefferson Lab. Unlike $B$- and $J/\psi$ decays
the photoproduction reaction has not been discussed from the
theoretical side. Therefore we shall choose it as a starting point
in order to introduce the effective range formalism. Then we shall
return  to Belle and BES data.

The double differential cross section for the reaction $\gamma p
\to pp\bar p$ is given by the well-known Chew-Low expression
\be
\frac{d^2\sigma}{ds_2 dt_1} = \frac{1}{2^{12}\pi^4}
\frac{\lambda^{1/2} (s_2, m^2, m^2)}{m^2k^2s_2} \int
d\Omega^*_{23}|T(s_2, t_1,\Omega^*_{23})|^2,\label{1} \ee where
$m$ is the nucleon mass, $k$ is the energy of $\gamma$,
\be
s_2=(p_2+p_3)^2,~~t_1= (k-p_1)^2,\label{2}\ee with $p_2$ and $p_3$
being the   4-momenta of $p$ and $\bar p$ forming the low-mass
pair, $p_1$ being the 4-momentum of the remaining proton,
$\lambda(x,y,z) = (x-y-z)^2 -4yz$ is the standard kinematical
function. The angle $\Omega^*_{23}= (\cos \theta, \varphi) $ is
defined in the CM system of the particles  (2,3). The variable
$s_2=m^2_{23}$ is the square of the invariant mass of the (2,3)
pair. It follows from (\ref{3}) that the angular distribution of
particles 2 and 3 ( the low-mass $p\bar p$ pair) is isotropic in
their CM system provided $T(s_2, t_1, \Omega^*_{23})$ is
independent of $\Omega^*_{23}$ which is the case for $^1S_0$ and
$^3P_0$ states. Equation (\ref{1}) yields the following invariant
mass distribution \be \frac{d\sigma}{dm_{23}} =
\frac{1}{2^{10}\pi^4} \frac{|\vep_1|}{km^2}
({m^2_{23}-4m^2})^{1/2}\int d(\cos\beta) \int d\Omega^*_{23}
|T|^2,\label{3}\ee where $\beta $ is the angle between $\vep_1$
and   the direction of  the incident $\gamma$ ( we remind that the
index 1 is attributed to the proton which remains outside the
correlated  $p\bar p$ pair).

In the spirit of the Migdal-Watson FSI theory we single out from
the matrix element $T$ a factor responsible for the low energy $p
\bar p$ interaction. Within Migdal-Watson approach one has
\be
|T|^2 = |T^{(0)}|^2/|f(-q)|^2 = D(q) |T^{(0)}|^2,\label{4}\ee
where $q$ is the $p\bar p$ CM momenta, $q^2=\frac14
(m_{23}^2-4m^2)$, $f(-q)$ is the Jost function corresponding to
the $p\bar p$ interaction at low energy, $D(q) = |f(-q)|^{-2}$ is
called the enhancement factor. As it was already mentioned the
dynamics of $N\bar N$ interaction  is much more complicated than
that of $NN$ \cite{8}, \cite{9}, \cite{11}, \cite{16}.
 Annihilation
dominates at short and possibly intermediate distances, $\omega$-
and other odd $G$-parity exchanges lead to a strong attraction to
be added to
 the one-pion exchange in the outer region \cite{17}. Therefore the
approximation of $f(-q)$ by a Born term from one-pion exchange
\cite{6} misses the essential features of $N\bar N$ dynamics. On
the other hand effective range solution for the low-energy $N\bar
N$ amplitude describes the whole set of data with a fair accuracy
\cite{12}-\cite{15}.

Ignoring for the moment complications due to annihilation, Coulomb
interaction and spin-isospin structure we may write the following
expression for Jost function in the scattering length
approximation
 \be
f(-q) \simeq A(q) (1-iqa),\label{5}\ee where $a$ is the scattering
length and  $A(q)$ is a well defined smooth function. A similar
expression for $p\bar p$ system is  not as simple for  three
reasons:

(i) $p\bar p$ system is a combination of the two isospin states
with $I=0,1$; isospin invariance is violated by the mass
difference of proton and neutron;

(ii) a powerful annihilation results in the complexity of the
scattering lengths;

(iii) the Coulomb attraction acts in the $p\bar p$ system.

Experimental data on $N\bar N$ spin observables exist only at the
incident $\bar p$ momenta $P_L> 500$ MeV/c \cite{18,8}, i.e.
beyond the energy region where the effective range approximation
is applicable. From the level  shifts in $p\bar p$ atom \cite{19}
it is possible to extract only some information on the imaginary
parts of the triplet and singlet  scattering lengths \cite{20}.
Therefore only spin-averaged effective range parameters were
extracted from the experimental data \cite{12}-\cite{15}.

 The enhancement factor in which all three points (i) -(iii)
 inherent for the $p\bar p$ system are accounted for has the
 following form \cite{12}-\cite{15}
 \be
 D(q) =\frac{c^2(q)}{|1-is (\tilde q
 +i\Delta  +l)-(\tilde q + i\Delta) lr|^2},\label{6}\ee
 where
 \be
 c^2(q) = \frac{2\pi}{qa_B}\left [ 1-\exp \left( -
 \frac{2\pi}{qa_B}\right) \right]^{-1}\label{7}\ee
 is the Sakharov Coulomb enhancement factor \cite{21} with $a_B =2/\alpha m =57.6$
 fm being the Bohr radius of the  $p\bar p$ atom,
 \be
 s=\frac12 (a_0+a_1), ~~ r=a_0a_1,\label{8}\ee
 where $a_I, I=0,1$ are $N\bar N$  $S$ -wave scattering  lengths
 with isospin $I$, \be
 \tilde q = c^2 (q) q +  \frac{2i}{a_B} h(qa_B),~~ h(z) =\ln z
 +Re\psi \left (1+\frac{i}{z}\right),\label{9}\ee
with  $\psi (z) =d/dz\ln \Gamma(z).$ The  quantity $l$ is the
 momentum in the $n\bar n$ channel,
\be
 l^2=q^2-m\delta,~~ \delta = 2(m_n-m),\label{10}\ee
 (recall that $m$ is the proton mass).

 The point $q=({m\delta})^{1/2} \simeq 49$ MeV/c corresponds to the
 $n\bar n$ threshold; below this point $n\bar n$ momentum $l$
 becomes imaginary, $l=i(m\delta-q^2)^{1/2}.$
Parameter $\Delta $ is the Schwinger correction to the scattering
length \cite{22}, $\Delta \simeq - 0.08$ fm \cite{12}-\cite{15}.

The enhancement factor (\ref{6}) does not include the effective
range term and the contribution from nonzero orbital momenta.
According to \cite{12}-\cite{15} the effective range term is of
minor importance up to $q\simeq 150$ MeV/c i.e. to
$Q=m_{23}-2m\simeq 20$ MeV. We  note  in passing that for a
multichannel system the effective range may be   even negative
or complex \cite{23,24}. The $P$-wave  comes into play  a little earlier
especially as far as the total cross section is concerned.
Inclusion of  $P$-wave brings about two problems. First, the
$P$-wave  scattering  length (it has a dimension of fm$^3$) is
very sensitive to the fitting procedure (see \cite{12}-\cite{15})
and the present set of the experimental  data do not warrant a
stable solution for the $l=1$ amplitude. The second problem is the
following. With $P$-wave included equations become rather
cumbersome and the number of parameters increases substantially.
This may cause unnecessary doubts in the  reliability of the
proposed approach. We plan to consider effective range terms and
$l>0$ amplitudes in the next  publication. This would allow to
analyze the $p\bar p$ correlation function in a wider energy
range.

The most remarkable result of all fits \cite{12}-\cite{15} for low
energy $N\bar N$ parameters is a  clear  dominance of the $S$-wave
with $I=0$ over that with $I=1$. The corresponding scattering
lengths read
\be
a_0 = (-1.2+ i0.9) {\rm ~fm},~~ a_1 = (-0.1+ i0.4) {\rm
~fm},\label{11}\ee
\be
a_0 = (-1.1+ i0.4) {\rm ~fm},~~ a_1 = (0.3+ i0.8) {\rm
~fm},\label{12}\ee where (\ref{11}) and (\ref{12}) are
respectively the results of \cite{13} and \cite{14}. The sign
convention in \cite{12}-\cite{15} is  $ k \cot \delta = +1/a$. We
see that the absolute value of  $|Rea_0|$  is much larger than
that of $|Re a_1 |$. This is completely in line with the
observation by BES of a strong low-mass effect in the decay
process $J/\psi\to \gamma p \bar p$ and the absence of a similar
structure in $J/\psi \to \pi^0 p\bar p $ \cite{3}.

The two solutions given by (\ref{11}) and (\ref{12}) are somewhat
different. This is due to several reasons described in
\cite{12}-\cite{15}. The main source of ambiguity is the lack of
the experimental data at very low energies.
The effective range analysis which resulted in the parameter
sets (\ref{11})-(\ref{12}) was performed at the end of eighties. 
In view of the LEAR shut down the experimental situation did not
drastically change since that time. The only substantial new
piece of information concerns the total and annihilation $\bar p p$ 
and $\bar n p$ cross sections down to $p_{Lab}=35$ MeV/c \cite{25,26}. 
However in the analysis of these data
\cite{27,28} use was made of the effective range solution \cite{29}
performed in 1988 which was later updated in 1991 \cite{14} partly
by the original authors. The parameters set (\ref{12}) is just this
updated version. The effective range solution \cite{29} fairly well fits
the low energy cross sections \cite{25}-\cite{28} which is an indication that
the same is true for our parameter set (\ref{12}). We urge our colleagues
who are in possession of the experimental data to persuade the
analysis whith the parameter sets \cite{13,14}. An independent
phenomenological analysis of the experimental results \cite{25} was
performed in \cite{30}. The authors extracted the imaginary part of
the Coulomb-distorted $p\bar p$ scattering length. This imaginary part
corresponds to the averaging over spin and isospin.  The $n\bar n$
threshold which lies just within the region under investigation
was not taken into account. This fact should be kept in mind
when comparing our parameter sets which the value presented in
\cite{30}. The result of \cite{30} is $Ima_{cs}=0.69$ fm while from 
(\ref{11}) we get $\frac{1}{2}Im(a_0+a_1)=0.65$ fm (the index in $a_{cs}$ 
means that this quantity still contains the
Coulomb correction) . Again we conclude that the ``new'' data \cite{25}
do not contradict the ``old'' solution (\ref{11}).

The negative sign of $Rea_0$ may be interpreted  as an indication
that the potential in this channel is either repulsive or on the
contrary attractive and  strong enough to produce a bound state
somewhere below  the threshold. However such a simple
interpretation is not obvious in strong annihilation regime. This
problem is beyond the scope of the present work.

The above arguments  based on the analysis of the reaction $\gamma
p\to  pp\bar p$ led  us to the following conclusion. Within the
framework of the FSI theory and within the region of the
applicability of the scattering length approximation the $p\bar p$
effective  mass distribution is governed by the  function \be
F(Q)=( m^2_{p\bar p} - 4 m^2)^{1/2} D(Q) \simeq2(mQ)^{1/2}
D(Q),\label{13} \ee where

\be
Q=m_{p\bar p} -2m = 2m \left\{
\left(1+\frac{q^2}{m^2}\right)^{1/2} -1\right\},\label{14}\ee and
$D(Q)$ is defined by (\ref{6}). The phase-space factor $Q^{1/2}$
is trivial and plays the role of the universal background. The
function $D(Q)$ reflects the the  main features of the $p\bar p$
dynamics.Due to Coulomb attraction it diverges at threshold since
$C^2(q)\to 2\pi/a_Bq$ as $q\to 0$. Thus at $Q=0$ the distribution
$2(mQ)^{1/2} D(Q)$ starts from a finite  value $4\pi/a_B$. At
$q=(m\delta)^{1/2}$ the factor $D(Q)$ has a cusp due to the
opening of the $n\bar n$ channel.

The above conclusions also hold true for the reactions $B^+\to K^+
p\bar p $\cite{1}, $\bar B^0 \to D^0 p \bar p $\cite{2} and
$J/\psi \to \gamma p \bar p$\cite{3}.

An remark is due at this point. The BES data on $J/\psi
\to \gamma p \bar p $ were analyzed making use of the Breit-Winger
formula \cite{3}. We wish to remind that in the vicinity of the
threshold the Breit-Wigner function can be reexpressed in terms of
the effective range parameters \cite{31}.

Our main results are presented   in Figs. 1-2. In Fig. 1 we
display the  enhancement factor $D(Q)$
 corresponding
to the parameters set (\ref{11}). In this figure we also plot the
experimental points obtained by BES collaboration and taken from
Ref.\cite{3}. The normalization of the experimental points is
chosen in such a way that in the region of $Q\simeq 0.02$ GeV they
are in rough agreement with the theoretical curves. The comparison
therefore may serve only for the orientation purposes and in order
to give an impetus for the future detailed analysis of Belle and
CLAS data as soon as these data are available.

The curve $D(Q)$ corresponding to the  parameter set (12) is not
shown since it goes very close to the presented in Fig.1. In Fig.
2 we illustrate the roles played by the Coulomb attraction and
annihilation. The solid curve in Fig.2 is the same as in Fig.1,
the dashed one corresponds to the Coulomb interaction switched
off, while the dotted curve -- to switching off the annihilation.

Our main conclusions may be summarized as follows. The
near-threshold proton-antiproton enhancement has been observed in
reactions  driven by different mechanisms. Therefore the  effect
may be due to $p\bar p$ FSI. In order to verify this assumption we
have described the FSI in terms of the low-energy $N\bar N$
parameters taking into account the isospin structure of the
amplitude and the Coulomb interaction. The factorization  of the
FSI factor from the total amplitude has been explicitly  performed
for the reaction $\gamma p \to p p \bar p$. In turns out that the
data of  BES  collaboration \cite{3} are fairly well described by
the effective range solution obtained long ago from the analysis
of the LEAR data. Further conclusions would become  possible as
soon as new experiment data from BES, Belle and CLAS are
available. Finally we remark that similar kind of analysis may be 
applied to $\Lambda \bar \Lambda$ final states. The low energy 
parameters for this system may be found e.g. in \cite{32}.

The authors are indebted to L.N.Bogdanova, A.Kudryavtsev, K.Mikhailov, 
P.Pakhlov, Yu.Simonov and B.S.Zou  for useful discussions and clarifying 
remarks. We are grateful to Dr. Shi-Lin for drawing our attantion to 
paper \cite{7}. B.K. acknowledge the support from the grant Ssc-1774-2003.

\newpage

% PICTURE 1
\begin{picture}(0,530)
\put(-100,0){\includegraphics[width=1.4\textwidth]{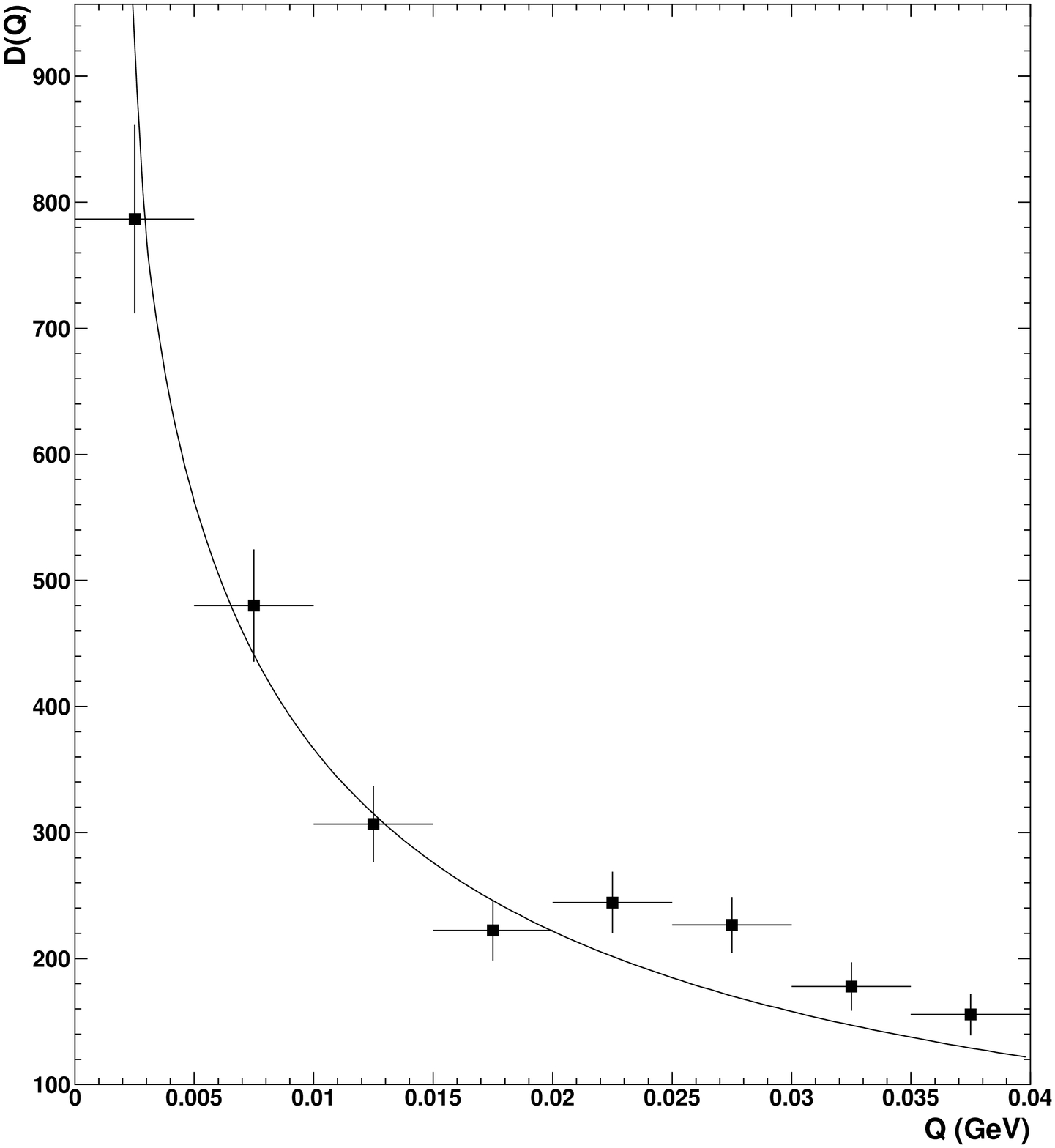}}

\put(-55,0){Fig. 1 The near-threshold  behaviour of the
enhancement function $D(Q)$, $Q= m_{23}-2m$.}
\put(-55,-15){Solid curve corresponds to the parameter set (\ref{11}).
Experimental points are from  Ref.\cite{3}.}

\end{picture}

\newpage

% PICTURE 2
\begin{picture}(0,530)
\put(-100,0){\includegraphics[width=1.4\textwidth]{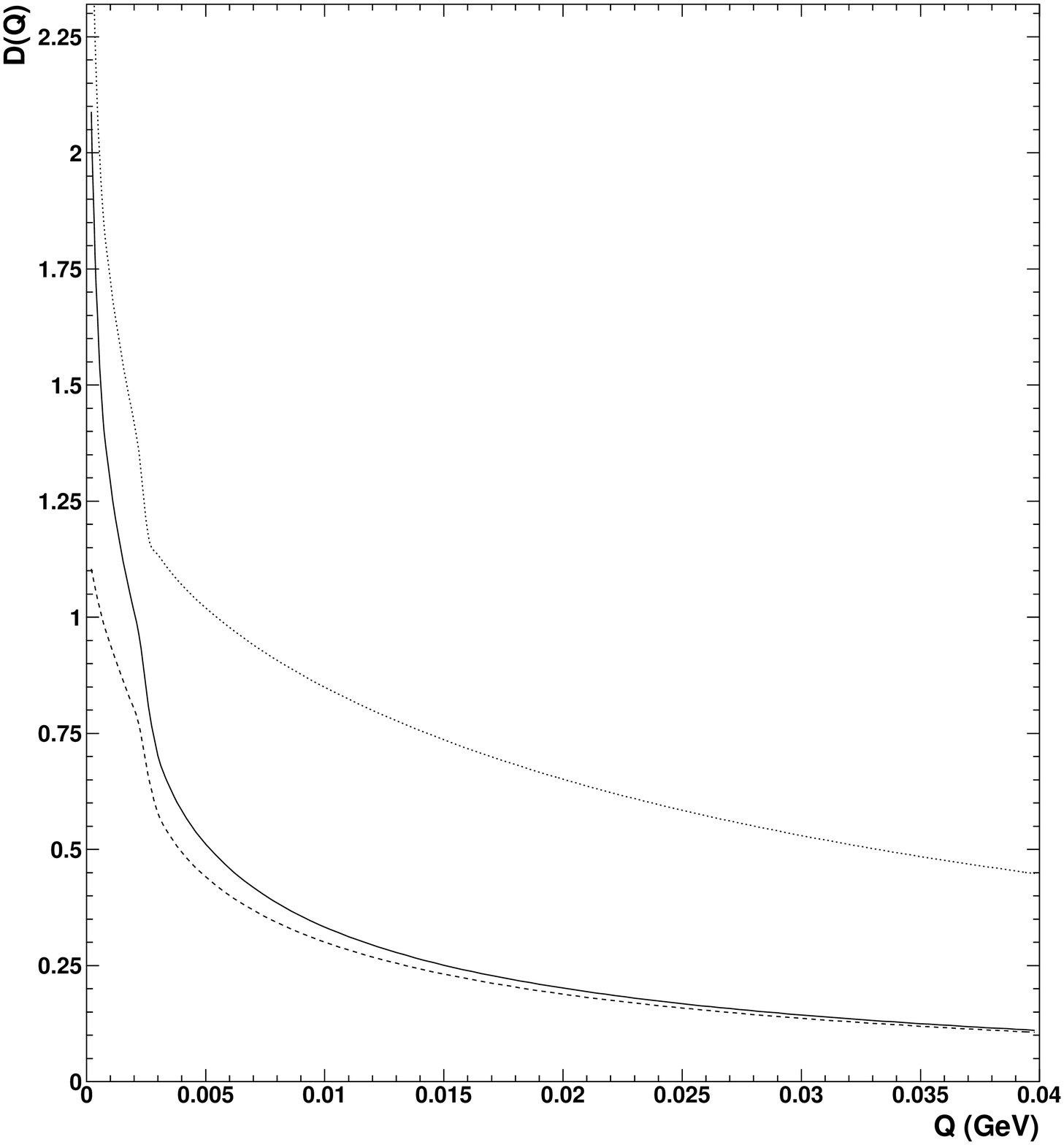}}

\put(-70,-5){Fig. 2   The enhancement factor according to (\ref{6}) with
parameter set (\ref{11}) -- the solid curve, the}
\put(-70,-20){same without Coulomb interaction -- the dashed curve, 
without annihilation -- the dotted curve.}

\end{picture}

\end{document}